\documentclass[nocopyright]{cimento}

\setcopyright{CERN on behalf the XXXXX Collaboration}
\usepackage{graphicx}  
\usepackage{booktabs}

\title{Optimization of the light detection system of the ICARUS detector}

\author{
  C.~Saia\from{ins:INAFCT}\ETC,
  C.~Petta\from{ins:INFNCT}\from{ins:UNICT},
  G.~L.~Raselli\from{ins:INFNPV} \atque
  M.~Rossella\from{ins:INFNPV}
  \\ on behalf of the ICARUS collaboration
}

\instlist{
  \inst{ins:INAFCT} INAF-OACT, Catania, Italy
  \inst{ins:INFNCT} INFN, Sezione di Catania - Catania, Italy
  \inst{ins:UNICT} Università degli Studi di Catania - Catania, Italy
  \inst{ins:INFNPV} INFN, Sezione di Pavia - Pavia, Italy
}

\begin{document}

\maketitle

\begin{abstract}
The ICARUS detector, a key component of the Short Baseline Neutrino (SBN) Program at Fermi National Acelerator Laboratory (FNAL), is a 600-ton Liquid Argon Time Projection Chamber (LArTPC) equipped with a Light Detection System (LDS) that uses 360 Hamamatsu R5912-MOD 8-inch photomultiplier tubes (PMTs), specifically designed to operate under cryogenic conditions ($\sim 87 \ K$). These PMTs feed the trigger signal to the readout, improve the spatial and timing resolution of the events, and contribute to cosmic rays mitigation.
During operation at FNAL, a progressive degradation in the PMT gain was observed. We developed an experimental setup to investigate the temperature dependence of PMT performance. Gain measurements were carried out from room temperature to $-70 ^\circ C$ using an environmental chamber. The results show that, while the PMTs exhibit stable performance at room temperature, a significant and irreversible reduction in gain emerges at lower temperatures. Although $-70 ^\circ C$ remains above the liquid argon temperatures, the trend clearly reveals a gain-sensitive degradation mechanism.
A simplified physical model was developed to reproduce and interpret the observed behavior. Based on these findings, a series of mitigation strategies were implemented in the ICARUS detector to preserve PMT performance and ensure reliable operation under cryogenic conditions.
\end{abstract}

\section{The Short-Baseline Neutrino (SBN) Program}
    SBN program was established in order to clarify the neutrino oscillations at the eV mass-scale scenario.
    It consists of two Liquid Argon Time Projection Chambers (LAr-TPCs): a Far and a Near Detector. They are respectively ICARUS and SBND and they are located at 600 m and 110 m from the Booster Neutrino Beam target~\cite{machado,proposal}. 
    Nevertheless, ICARUS is located $6°$ off-axis of the Neutrinos at the Main Injector (NuMI) beam, that will be used to perform detailed studies of neutrino-argon interaction cross-sections~\cite{Abratenko}. 
    The ICARUS detector has been collecting data starting with RUN1 in June 2022, and RUN5 data taking is going on.

\section{ICARUS T600 and its Light Detection System} \label{sec1} 
    ICARUS detector consists of two identical T300 modules filled with liquid Argon. In each module, electrons produced by ionizing particles drift from a common central cathode towards the anodes, which consist of three parallel planes of wires.
    These ionizing particles also produce scintillation light in the VUV range ($\lambda$~=~128~nm), collected by 360 PMTs installed behind the planes of wires. The PMTs feed the trigger signal to the readout, improve the spatial and timing resolution of the events, and contribute to cosmic rays mitigation.
    In order to guarantee the detector operation at shallow depth a Cosmic Ray Tagger (CRT) system and a 2.85~m concrete overburden were installed~\cite{proposal,LDS}.

\section{The Photomultiplier}
\label{subsec2}
    A set of 360 PMTs (8'' Hamamatsu R5912-MOD) has been characterized prior to installation. They have a bialkali photocathode on a platinum under-layer, allowing operation at temperatures as low as -200°C. Their sensitivity is peaked around 420 nm. VUV photons are wavelength shifted to visible light with 200 $\mu$g/cm² of Tetraphenyl Butadiene (TPB). The typical gain is $10^7$ at 1500 V, with a linear anodic response to the operating light intensity~\cite{calibration}. 

\section{Observation at FNAL}
Since the start of operations in 2021, the PMTs, operating at the cryogenic temperature of liquid argon (87 K), exhibited a significant and progressive gain reduction. Initial measurements showed a loss rate of approximately 1.93\% per month.

It was hypothesized that the gain reduction was due to a decrease in the intrinsic gain of the PMT, originating within the dynode amplification chain, rather than a deterioration of the photocathode's quantum efficiency. Measurements of the rate of background optical pulses over time showed no decrease, indicating that the photocathode continued to convert photons into photoelectrons at the same rate. 

Maintaining stable PMT gain is crucial for ICARUS because it ensures precise timing, accurate event localization within the detector, and reliable trigger performance to distinguish neutrino events from cosmic-ray background. A significant and uncontrolled reduction of the PMT gain would directly impact the performance of the ICARUS detector. Weak signals would make it harder to distinguish real events from electronic noise, potentially lowering the trigger efficiency and causing some neutrino interactions to be missed. Time resolution would also worsen, reducing the accuracy in determining the event t0, which is crucial for 3D track reconstruction and cosmic ray rejection.

\section{Mitigation Strategies Implemented in ICARUS}
To counteract the gain loss and ensure the long-term stability of the detector, several mitigation strategies were successfully implemented:
\begin{itemize}
    \item \textit{Overburden Installation} (June 2022): The installation of the concrete overburden reduced the cosmic ray flux, thereby decreasing the amount of background light hitting the PMTs and slowing the degradation rate.
    \item \textit{Gain Reduction}: The operating gain of the PMTs was lowered. During RUN1 and RUN2, the average gain was reduced from $0.7\times10^7$ to $0.46\times10^7$. This reduced the stress on the dynodes and decreased the monthly loss rate from 1.93\% to 0.64\%.
    \item \textit{New Signal Cables}: Before RUN3, new, higher-performance signal cables were installed. This improved the signal shape and amplitude, allowing for a further reduction in the operating gain to $0.39\times10^7$ while maintaining excellent signal quality. This final step lowered the gain loss rate to $0.31\%$ per month.
\end{itemize}

\section{Experimental Study at INFN Catania}
To investigate the phenomenon in a controlled environment, an experimental setup was developed at INFN Catania~\cite{Saia}.

\begin{figure}[htb]
    \centering
    \includegraphics[width=\linewidth]{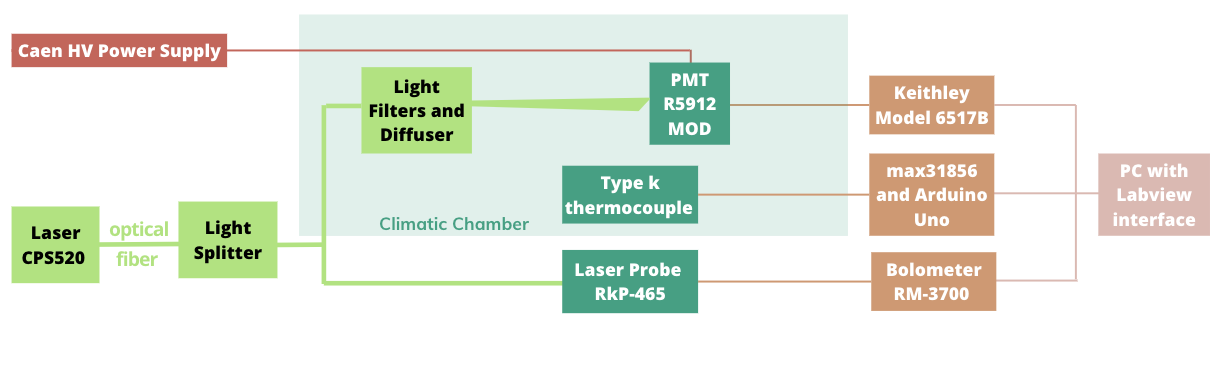}
    \caption{\textit{Schematic of the experimental setup}}\label{schema}
\end{figure}

A single PMT without TPB coating was placed in an environmental chamber with controlled temperature down to -70°C.
The PMT was illuminated by a continuous laser and operated in current mode to simulate the high-charge conditions at FNAL. The measurement is performed in current mode, stressing the PMT with the maximum anodic current tolerated by the PMT electronic base ($\sim 28 \ \mu A$)~\cite{Saia}. 

As shown in Fig. \ref{schema}, a laser beam ($\lambda$ = 520 nm) is split into two optical fibers: the first beam is attenuated and then diffused over the surface of the PMT photocathode.
The other optical fiber is used for an independent reference measurement of the injected light power via a bolometric photodetector. The temperature is measured by a type K thermocouple located near the PMT photocathode. A DAQ system recorded the PMT's anode current, a reference light intensity, and the temperature~\cite{Saia}.

\section{Results}
The performance of the photomultipliers at different temperatures was studied by monitoring the \emph{anode current}, corrected by light intensity fluctuations (less than $1\%$). We carried out two sets of measurements with different illumination conditions. As mentioned before, the observed reduction in anode current is interpreted as a loss in gain, since a decrease in the photocathode quantum efficiency has been excluded as a possible cause~\cite{Saia}.

\begin{figure}[htb]
  \centering
  \includegraphics[width=0.7\textwidth]{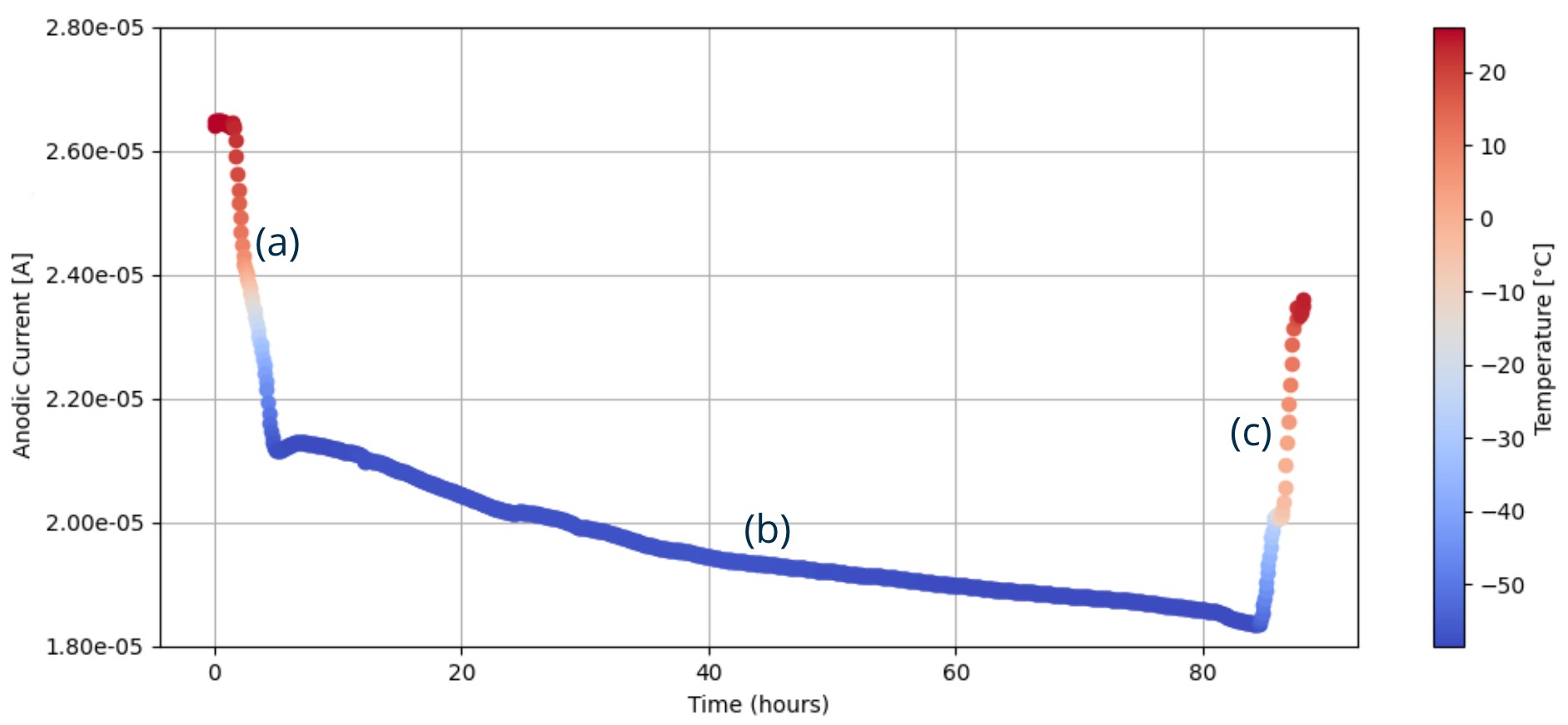}
  \caption{Example of anode current over time: reversible reduction and irreversible degradation at low temperature~\cite{Saia}.}
  \label{fig:cooling}
\end{figure}

\begin{figure}[h!]
    \centering
    \includegraphics[width=0.7\linewidth]{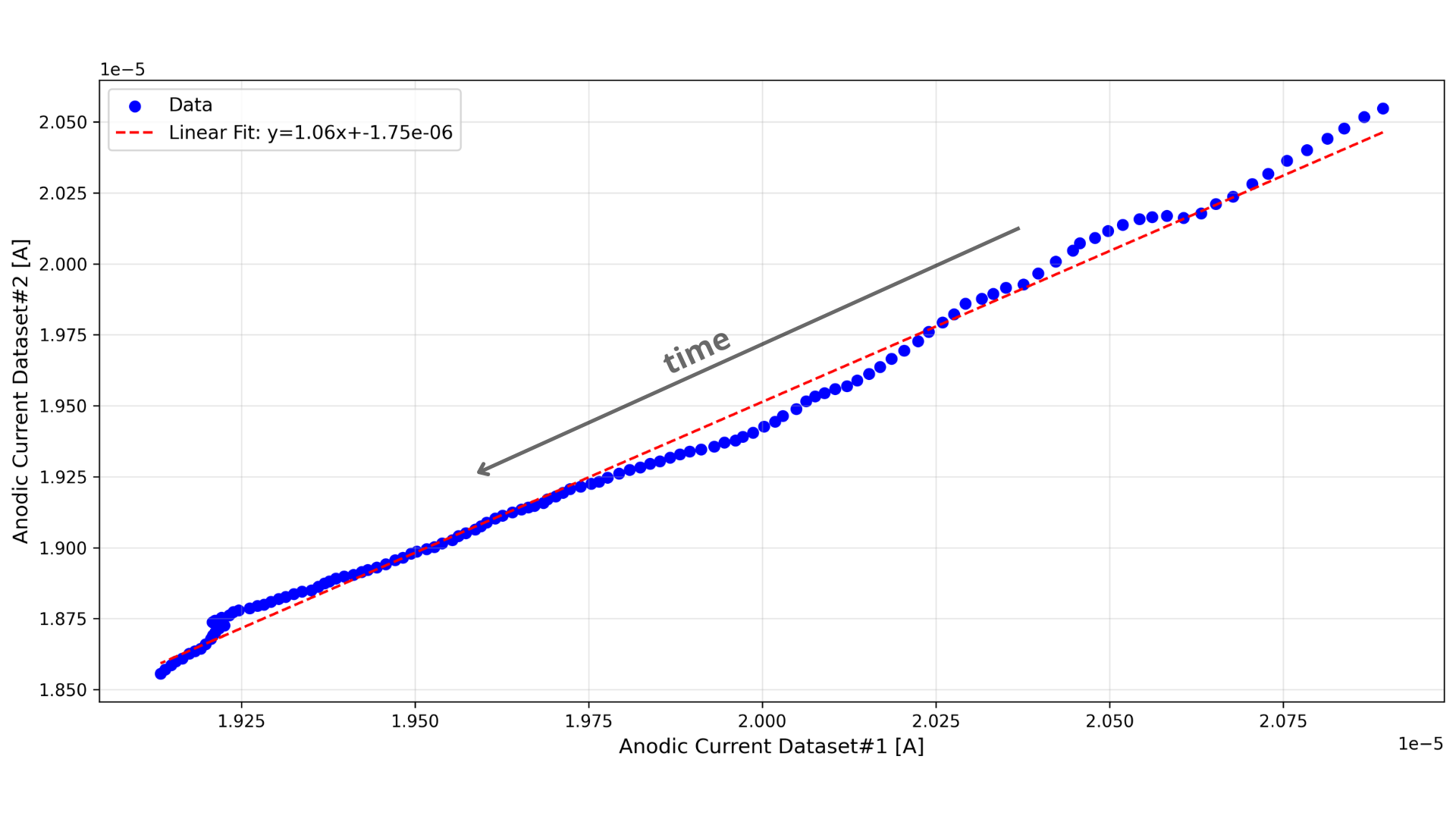}
    \caption{\textit{Time correlation of two datasets at the same temperature and sharing the same initial anode current}}
    \label{fig:confronto2}
\end{figure}

\begin{itemize}
    \item \textbf{Room Temperature Stability:} no gain loss was observed when the PMT was operated for extended periods at room temperature, even under high current.
    \item \textbf{Low Temperature Degradation:} during cooling cycles to -60/-70°C, a clear and permanent gain loss was observed. 
    The observed behavior can be summarized in three main points, illustrated in Fig.~\ref{fig:cooling} as (a), (b) and (c). When the temperature is lowered, the anode current decreases (a). This effect can be attributed both to variations in the quantum efficiency (q.e.) and to a reduction of the effective gain. Importantly, this behavior is \emph{reversible}: once the device is brought back to room temperature, the current would return to its initial value if no other effects occur (c)~\cite{Saia}.  
    
    In contrast, we observe a degradation that is not reversible.  Even after restoring the operating temperature to room conditions, the current does not recover its initial value, indicating a permanent damage (b).

    \item \textbf{Reproducibility:} 
    It is interesting to note that measurements performed under different illumination and voltage conditions, but sharing the same initial anode current, exhibit very similar trends. 
    The trend, as shown in the plot in Fig. \ref{fig:confronto2}, clearly indicates a high degree of correlation between the two datasets, despite the different operational conditions. This suggests that the observed behavior is primarily driven by the anode current rather than by individual dependence from light intensity or applied voltage.

\end{itemize}

\section{A Simple Model}
Based on the considerations made so far, we developed a simplified physical model to reproduce and interpret the behavior. The secondary electron emission ratio $\delta_i$ for the i-th dynode is proportional to the interstage voltage applied $V_i$. The PMT gain can be written:
\begin{equation}
        G = \prod_{i=1}^n \delta_i = \prod_{i=1}^n A_i V_i^{\alpha_i} = \prod_{i=1}^n A_i (\epsilon_i V)^{\alpha_i}
\end{equation}

The idea is that all dynodes have the same initial conditions but over time they can suffer a degradation in the secondary emission coefficient. We can reasonably hypothesize that the degradation of the dynodes depends on the amount of current at their stage. However, only the last dynodes in the chain should impact on the total gain loss, since the first dynodes receive much less current and therefore degrade less. This degradation alters the parameters of the secondary emission coefficient, specifically $A$ or $\alpha$.

In Model A, we assume the coefficient $A$ to decrease proportionally to the current in each multiplication stage. This loss factor acts every time an electron strikes a dynode and is applied individually for each dynode over various time steps. A similar loss factor was also introduced for the coefficient $\alpha$ in Model B. The simulation uses the known voltage divider data to model the dynode multiplication chain \cite{pmt}.  

\begin{figure}[htb]
    \centering
    \includegraphics[width=0.9\linewidth]{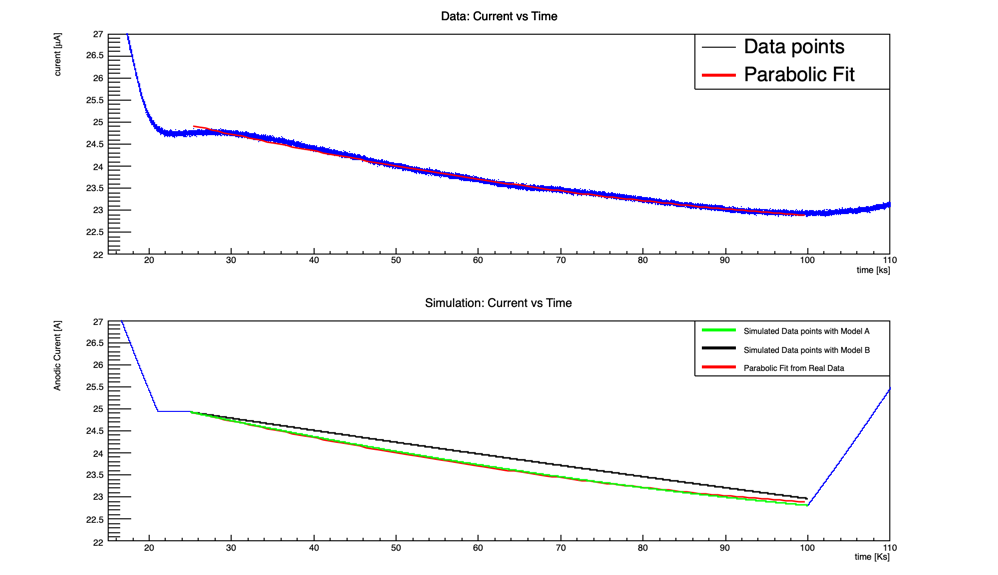}
    \caption{\textit{Top: Experimental data from the first cooling phase, fitted with a parabolic fit. Bottom: Comparison between two degradation models. The fit from the top graph is included for reference.}}
    \label{fig:AAAAAAA}
\end{figure}

As shown in Fig.~\ref{fig:AAAAAAA} (bottom), Model A -- i.e. a loss factor proportional to the current affecting the parameter $A$ -- better reproduces the observed degradation trend. This seems reasonable because $\alpha$ depends on the material of the dynodes, which we do not expect to change significantly at low temperatures. Instead, in multilayer dynodes, mismatches in the thermal expansion coefficients of different materials can induce significant mechanical stress at low temperatures, potentially leading to microcracks or delamination when several accelerated electrons impinge on them. This surface degradation can be expressed as a decrease in the factor $A$  \cite{thelastcit}. 

\section{Conclusions}
The studies performed at INFN Catania confirm that the gain loss observed in laboratory conditions reproduces, at least qualitatively, the behavior seen in ICARUS during operation at Fermilab. Even if the environmental chamber tests were carried out at temperatures higher than liquid argon, the results clearly point to a current-dependent degradation mechanism affecting the last dynodes of the PMTs.

Most importantly, the mitigation strategies implemented in ICARUS --- reduction of the operating gain, installation of the overburden, and replacement of the signal cables --- have proven effective in stabilizing the PMT response. Thanks to these optimizations, the Light Detection System ensures a reliable trigger and accurate timing information, allowing ICARUS to perform long-term neutrino measurements with the required sensitivity.

In conclusion, the detector is now able to sustain stable and robust operation throughout the expected lifetime of the Short-Baseline Neutrino Program, ensuring the physics goals of ICARUS can be fully achieved.

\acknowledgments
This manuscript has been authored by Fermi Forward Discovery Group, LLC under Contract No. 89243024CSC000002 with the U.S. Department of Energy, Office of Science, Office of High Energy Physics.

\end{document}